# A Conversation With Harry Martz

**Paul Kvam**

*Abstract.* Harry F. Martz was born June 16, 1942 and grew up in Cumberland, Maryland. He received a Bachelor of Science degree in mathematics (with a minor in physics) from Frostburg State University in 1964, and earned a Ph.D. in statistics at Virginia Polytechnic Institute and State University in 1968. He started his statistics career at Texas Tech University's Department of Industrial Engineering and Statistics right after graduation. In 1978, he joined the technical staff at Los Alamos National Laboratory (LANL) in Los Alamos, New Mexico after first working as Full Professor in the Department of Industrial Engineering at Utah State University in the fall of 1977. He has had a prolific 23-year career with the statistics group at LANL; over the course of his career, Martz has published over 80 research papers in books and refereed journals, one book (with co-author Ray Waller), and has four patents associated with his work at LANL. He is a fellow of the American Statistical Association and has received numerous awards, including the *Technometrics* Frank Wilcoxon Prize for Best Applications Paper (1996), Los Alamos National Laboratory Achievement Award (1998), R&D 100 Award by *R&D Magazine* (2003), Council for Chemical Research Collaboration Success Award (2004), and Los Alamos National Laboratory's Distinguished Licensing Award (2004). Since retiring as a Technical Staff member at LANL in 2001, he has worked as a LANL Laboratory Associate.

These conversations took place by phone and e-mail during the summer and fall of 2005.

## 1. EARLY LIFE

**Kvam:** Harry, I'm glad to have known you now for 15 years, and I am even more pleased to interview you for *Statistical Science*. It's quite an honor for you and a fitting cap to your remarkable career in statistics. I'd like to start by asking about your childhood. What was it like growing up in rural Maryland in the 1940s and 1950s?

*Paul Kvam is Professor, School of Industrial and Systems Engineering, Georgia Institute of Technology, Atlanta, Georgia 30332-0205, USA e-mail: paul.kvam@isye.gatech.edu.*

**Martz:** The 1940s and 1950s in western Maryland, as in all America, was a far simpler era. Life certainly seemed slower and less hectic then, with far fewer demands on my time. It was life in the slow lane. I enjoyed a very happy childhood playing cowboys and Indians, shooting basketball, hunting, fishing, camping, hiking, and building model cars, boats and airplanes. I was blessed to have several really good friends (particularly Danny Moreland, Bill Claus, Wally Harshberger and Lynn Dehart), and we still keep in touch. I loved school, made good grades, and was usually in the top section of my class. When I was perhaps 10 years old or so, I somehow happened on an account of the Manhattan Project in Los Alamos and was fascinated by the development of the "bomb" there. Little did I know then that I was destined to spend most of my career in Los Alamos.

**Kvam:** Can you tell me more about your mother and father?

**Martz:** I believed then, and still do, that I had the best parents in the world. My father was a simple, hard-working, honest, intelligent, and kind man of German ancestry. He was completely devoted to my mother and me. He was extraordinarily skilled with his hands and loved to build things, especially with plans found in *Popular Mechanics* or *Mechanix Illustrated*. He was my best friend and always made time for me. He was especially compassionate, loving, and devoted. Although he had only an eighth-grade education, he had a Ph.D. in love, kindness, compassion, and generosity. He was a blue-collar shift worker employed by Potomac Edison Company as a control room operator at a local coal-fired power plant.

My father also loved solving puzzles of all kinds. He had the talent and ability to make small wooden and metal puzzles and was always challenging me to solve his latest creation, which I usually failed to do, generally to his delight.

My mother, who also was of German extraction, was a wonderfully warm stay-at-home mom, whose interests centered on cooking, keeping a clean and tidy home, and her family and friends. On numerous occasions I remember taking a pot of leftover soup to "Mrs. Lewis," who was a poor, but kind, elderly widowed neighbor. My mom lived The Golden Rule and probably never knew how much she taught me by example. She demonstrated Christian love rather than talking a whole lot about it.

**Kvam:** What were your teenage years like?

**Martz:** As I said before, growing up in the 1950s was wonderfully simple, as that time in America was an innocent age when Christian moral values were clearly taught and reinforced by parents, schools, and the church. The Ten Commandments applied to everyone, and freedom of religion meant that I was free to attend the Christian church of my choice. How very different in America today! Church was an important part of my life and, because I am a good speaker, I was often called upon to deliver the sermon on Youth Sundays. Youth fellowship at the church was an integral part of my teen years, and that, along with Boy Scouts, played an important role in teaching me vital life skills and moral values.

**Kvam:** What did you do for fun?

**Martz:** Lots of things! In those days fun didn't come from the latest computer game, and I often had to "work and invest" in having fun. I was and still am an avid reader. I remember reading the entire Dave Dawson War Adventure Series in just a few weeks one summer. Weekday evenings from 5:00 to 6:00 p.m. would find me listening to "The Lone Ranger" and "Sky King" on the radio; on Saturday mornings, it was "Buster Brown" and "Big John and Sparky." As a child I was also encouraged to play, and play I did! I built and flew model control-line airplanes; assembled innumerable plastic model cars, boats, and airplanes; fashioned all sorts of mechanized gizmos from my Lionel erector set; made numerous toys out of wood; built a "hut" in the woods behind our house; camped out in an Army pup tent in the backyard; skied to Bill Claus' house on a nearby hill; and "plinked" a lot with an old falling block single-shot 22 rifle.

**Kvam:** So your early interests were more pointed to science and engineering than, say, sports or the arts?

**Martz:** I dabbled in football in the ninth and tenth grades but quickly realized that, at $5'7''$, it was a fool's pursuit. I was also active in boys Hi-Y, band, orchestra and the National Honor Society. I played a trumpet in both marching band as well as a big-band dance band. I also played first-chair French horn in orchestra. In about the ninth grade, I decided to major in engineering in college, most likely mechanical engineering, because I am good at math and good at designing and building things. I planned to live at home while enrolling in the preengineering program at Frostburg State University, which was only six miles from my home, for two years and then planned to transfer to the University of Maryland for the remaining two years. However, after two years at Frostburg, I liked mathematics and science so much that I decided to stay at Frostburg and major in mathematics and minor in physics. It was much cheaper than transferring to the University of Maryland, as we didn't save much for college expenses.

**Kvam:** What caused this interest in mathematics and physics?

**Martz:** I realized early in school that God gave me talents for math and science. I liked the objectivity of math in contrast to, say, English or history, which seemed to me to be quite subjective. With math you either solved a problem correctly or you didn't, and I liked the assuredness and finality of that. Science appeared to me to be basically a lot of answered and unanswered questions forming a gigantic mind-numbing puzzle. And how I love puzzles! Physics is concerned with why and how the physical universe works, coupling nicely with my mechanical interests.

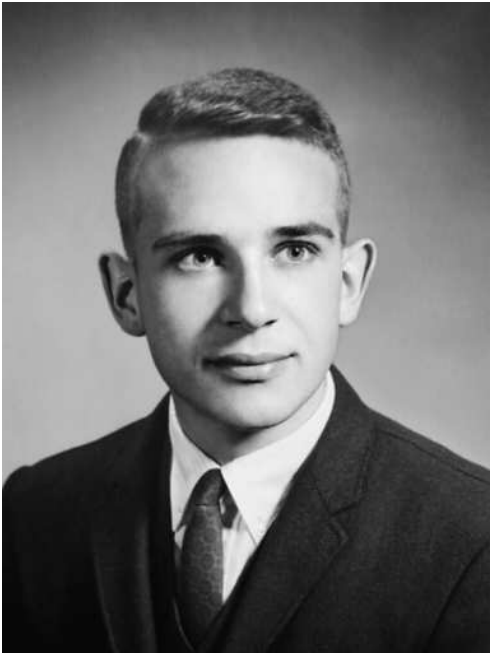

FIG. 1. *Harry Martz in 1960 as a senior at Allegany High School, Cumberland, Maryland.*

**Kvam:** How did you then become interested in statistics?

**Martz:** During my junior year at Frostburg I decided to pursue a Ph.D. in either mathematics or some closely related field. I took only a single-semester statistics course while at Frostburg using Brunk's 1960 book *An Introduction to Mathematical Statistics* (Brunk, 1960). Despite receiving a grade of "B," I enjoyed the course. I liked the obvious link to real-world problems through data analysis and could clearly see the practical utility of statistics. I even liked the permuted backward (inductive) thinking required, although this process was clearly different from other math courses. Thus, on a "lark," I subsequently decided that I would at least apply somewhere to a graduate statistics program. To make a long story short, near the end of my senior year I was offered a teaching assistantship in the Math Departments at both the University of Illinois and the University of Maryland, but was offered a National Institutes of Health fellowship in the Statistics Department of Virginia Polytechnic Institute and State University. I'm smart enough to realize that "free" money is better than "earned" money, so I entered Virginia Tech in the fall of 1964 to study statistics.

## 2. AFTER COLLEGE

**Kvam:** What was your graduate study like?

**Martz:** Virginia Tech had a large graduate program at the time that was quite competitive. And I thrive on competition! In those days, all grades were posted by student name by the classroom door—how different today. I did very well and, subsequently, completed all the requirements for a Ph.D. degree in statistics [with a minor in Operations Research (OR)] in August 1967. My dissertation was on empirical Bayes and was directed by Dr. Richard Krutchkoff. Dick was a student of Herbert Robbins while at Columbia University, and so I guess this makes me Herb's academic grandson.

**Kvam:** You were married in college?

**Martz:** Early in my first year at Virginia Tech, I married Rosalie Scimonelli, whom I met and fell in love with during my senior year at Frostburg. We were subsequently blessed by the birth of a delightfully precocious son, Joseph Christopher, in June 1965. We spent many a Saturday night playing bridge together with our friends John and Natalie Cornell while our children slept or played nearby. John was also a statistics graduate student at the time. Those days were extremely busy, and on most weekday evenings I would be in my carrel at the library studying until after midnight. As I was to find out later, this pattern of behavior took a definite toll on my marriage.

## 3. ACADEMIC CAREER

**Kvam:** What did you do after you graduated?

**Martz:** I was pursued by and selected for interview trips to the OR department at Johns Hopkins, Bell Laboratories and the Industrial Engineering (IE) Department at Texas Tech University. However, after my first trip to Texas Tech, they needed an answer ASAP, and I had to say "yea" or "nay" before the other interview trips. Having flown into Lubbock, TX, I was totally unaware of its isolated location and just how large Texas really is. The university was large and very nice, and the IE department was growing under the very capable leadership of Dr. Richard Dudek. He was a very persuasive guy and I finally said "yes" to his offer to come to the IE department at Texas Tech.

**Kvam:** How did you like being in an IE department?

**Martz:** I liked the challenges in serving an engineering community as resident statistician. I have always believed that, if possible, statisticians should leave the security of their own kind and reside in academic locations where practitioners live. I never did believe that what I now jokingly refer to as "academic incest" is the best way to advance our craft. I taught such courses as engineering statistics, stochastic processes, quality control, inventory control, operations research, queuing theory, reliability and statistical computation. I also served as the principal investigator on several grants and directed the research of many students.

**Kvam:** Is that when you became interested in reliability research?

**Martz:** Yes, although I had not previously formally studied reliability, teaching it set the stage for a career-long interest in it.

**Kvam:** What were the hot topics of reliability research when you started in the 1960s?

**Martz:** In the reliability inference area, there was lots of interest in frequentist notions such as the best way to compute confidence intervals for series and other systems. There was also a great deal of interest in logical modeling efforts involving fault and event trees. Barlow and Proschan were hard at work developing coherent systems and reliability characterization theories such as classes of life distributions based on aging and more generally placing many new notions in reliability on a firm probabilistic foundation. That decade was also a heady time for reliability applications involving new ideas of system effectiveness, including maintainability and availability. Simple stochastic process models in reliability were just beginning to appear, and Bayesian reliability applications were merely a novelty.

**Kvam:** Did you have any Ph.D. students that made a splash in the scientific community?

**Martz:** Yes. One in particular comes to mind. My second Ph.D. student, G. Kemble Bennett, is presently Vice Chancellor and Dean of the College of Engineering as well as the Director of the Texas Engineering Experiment Station at Texas A&M University. Several of my other Ph.D. students have gone on to become department chairs at various IE departments throughout the country. Many of my former students are also now Fellows of various professional societies. One is now a minister.

**Kvam:** You were at Texas Tech for almost a decade. Why did you leave?

**Martz:** That period of my life was especially tumultuous. My second son, Jeffrey, was born in January 1969, and my daughter, Angela, arrived in 1972. Two truly wonderful joys in my life. Unfortunately, my marriage to Rosalie began to founder in 1971 and ultimately failed in 1975. It was not my choice, but what could I do? I knew at the time that this decision would be extremely devastating to my children. To recover from this tragedy and to regain my self-worth, during 1975 and 1976, I took a leave of absence from Texas Tech and joined Los Alamos National Laboratory as a Technical Staff Member in the statistics group.

## 4. LOS ALAMOS

**Kvam:** So then you started your career at Los Alamos National Laboratory. What was the statistics group at Los Alamos like when you started? Who was there?

**Martz:** It was rather small but with some really outstanding people who knew how to apply statistics. The group leader was Ron Lohrding and his deputy was Ray Waller, an energetic twosome who really knew how to get the best out of their staff. Dick Beckman, Mike McKay, Kathy Campbell, Larry Bruckner, Tom Bement and Gary Tietjen were there as well. It was during this period that McKay and Beckman, along with Jay Conover, produced their famous work on Latin hypercube sampling (McKay, Beckman and Conover, 1979).

**Kvam:** What was the atmosphere like in Los Alamos during those years?

**Martz:** It was generally quite pleasant, laid back and productive. Funding was easy to get, and we all enjoyed far less micromanagement than we experience today. It was tons of fun working with some of the nation's brightest and best scientific minds. The projects to which we were assigned consisted of multidisciplinary teams, and we were often required to develop some new twist to or embellish an existing statistical method. More often than not, this strategy resulted in an opportunity to publish the results in either the subject matter or an applied statistics journal. On classified projects, we weren't permitted to publish; however, we still felt a nice sense of accomplishment knowing that our work was important to national defense. In addition to our project assignments, we were also encouraged to pursue research of our own choosing, and there was funding available to do this. It was a nice blend of project commitments and independent research.

**Kvam:** Tell me about your role as the leader of the statistics group.

**Martz:** I was named group leader of the statistics group in November 1984 and stepped down to continue research in August 1988. Tom Bement was my deputy. Except for several thorny personnel issues, it was generally fun. I am particularly good at getting funding; thus, money was never a problem. I also hired several good people, and the group grew in size. Because I believe that research is truly "seed corn" for the future, I emphasized independent (nonproject-related) research and obtained necessary funding to allow the staff to do this. I was asked to take a two-year leave of absence from the statistics group in September 1989 to assume the duties as technical team leader of risk analysis/reliability in the newly formed New Production Reactor Safety Project Office. I did so, after which I then returned to statistics in September 1991 and worked there as a staff member until retiring in June 2001.

**Kvam:** What was the most interesting project you worked on?

**Martz:** This is a no brainer! Starting in the mid-1990s, I began working with reliability engineers (Tom Lange, Art Koehler and many others) at Procter and Gamble (P&G) under a government-sponsored cooperative research and development agreement (GRADA) to statistically model the reliability of some extremely complicated production processes used to manufacture consumer paper products. By their own admission, P&G was then "data rich but knowledge poor," meaning that they routinely collected loads of automated reliability-related data from their processes but, because they were overwhelmed by the sheer volume of data, they didn't know what useful information the data contained. That's where I (and later we) came in. Their manufacturing processes literally had hundreds of failure modes, all of which they wanted to simultaneously model. Together we developed reliability models that were able to do this, and they were even more successful than we could ever have imagined.

According to P&G, these methods have increased plant productivity up to 44 percent, cut controllable costs up to 33 percent, improved equipment between 30 and 40 percent, reduced the time for line change-overs from as much as one hour to as little as six minutes, achieved 60 to 70 percent faster new equipment and product start-ups, and saved P&G over $2 billion since implementation. P&G is presently using the methodology at all 200 of its plants worldwide.

We received several nice awards for this joint work: a R&D 100 Award from *R&D Magazine* in 2003, a 2004 Council for Chemical Research Collaboration Success Award, a Los Alamos 2004 Distinguished

Licensing Award, a 2006 Federal Laboratory Consortium Award, two patents, and signed one of the largest licensing agreements that Los Alamos National Laboratory ever signed with industry. The corresponding software, known as PowerFactor, is now being used by several other companies. Much of the recent work was done in collaboration with Mike Hamada, an extremely talented staff member here at Los Alamos. In the broadest sense, this GRADA is a real success story for statistics.

**Kvam:** Tell me about the book you wrote with Ray Waller (Martz and Waller, 1982). What gave you the idea to write it?

**Martz:** Ray and I began developing Bayesian reliability methods for the NRC (US Nuclear Regulatory Commission) in the late 1970s and wrote several technical reports and handbooks describing these techniques. We were also asked to prepare and teach several corresponding two- to three-day short courses for them as well. In the early 1970s, the NRC recognized the value of Bayesian methods in conducting probabilistic risk analyses of commercial nuclear power plants. About the same time, we recognized that there were no Bayesian reliability textbooks on the market; so we decided to write one, as we had already prepared much of the material—or so we thought—in the set of reports.

We needed funding and were able to talk the Advanced Reactor Programs Office of the U.S. Department of Energy (DOE) into supporting this effort. Consequently, I spent about half of my time at work during the next few years writing the book; the other half I spent on NRC and some other projects. Because of group office duties, Ray worked on the book primarily during evenings and weekends. As I recall, it took us slightly over two years to complete.

In 1982 I also met my wonderful wife, Carol Ann, while we were both teaching Sunday school at the First United Methodist Church in Los Alamos. After a brief courtship, we were married in November 1982—shortly after the book appeared. We were truly "The Brady Bunch" as we parented all six of our children together, although one was already in college when we married.

**Kvam:** What were your major research interests while at Los Alamos?

**Martz:** My primary interests here at Los Alamos largely centered on applications of Bayesian statistics. The main areas that I've worked in include reliability, health physics, probabilistic risk assessment, weapon systems and large-scale systems. I've also done research in genetic algorithms, complex manufacturing systems, knowledge elicitation and data combination and integration. My most recent interests include Bayesian networks. For reasons unknown to me, I've always been keenly interested in Bayesian reliability demonstration testing, probably because I enjoy the inductive thinking required in determining testing requirements. At Texas Tech University, I also did research in discrete-time linear systems involving Kalman filter theory.

**Kvam:** Anyone who's seen you give a talk knows that you have this unbridled enthusiasm for what you are doing. Where do you get that?

**Martz:** Good question—I really don't know. I do know that I love to give talks and generally like to "perform" for an audience. I guess that I've just learned what not to do having suffered through far too many technical talks in which I was simply like a dog watching television. Didn't understand a thing. Nada. Nothing. I have never believed that, by hiding in the mathematical bushes, a presenter has an appropriate excuse for a poorly delivered talk.

**Kvam:** You are also known as an intense worker. I was told that you once wrote an entire paper in one day. Is that true?

**Martz:** No. Actually, it took a weekend. As I recall, I had the idea for a new definition of availability (Martz, 1971), which I called single cycle availability, on a Friday. I worked on it over the weekend and sent in the short paper the following Monday or Tuesday.

**Kvam:** You're known as a "dyed-in-the-wool" Bayesian. This must have ingratiated you with other Bayesians, but did it also strain your relationship with the non-Bayesian community?

**Martz:** I've never really had much of a relationship with the frequentist community. My earliest work was on Kalman filtering, which is a Bayesian technique. Most of my later NRC-sponsored research was in probabilistic risk assessment, which is likewise built on Bayesian methods. Later, I found out that health physics is a natural area for applying Bayesian techniques. Because I worked with all kinds of engineers while in the IE Department at Texas Tech, they could not have cared less about the frequentist/Bayesian controversy. Engineers take to Bayesian methods like bears to honey. The professional jabs and squabbling between the two camps were simply an odd amusement to them.

**Kvam:** I don't think I want to pursue that analogy by asking you who the "bees" are. Where does Bayesian methodology stand in reliability today, and where do you see it going?

**Martz:** Well, I firmly believe that Bayesian methods in reliability will continue to be developed and applied for two compelling reasons. First, they are just so very powerful for accommodating all the nuances that surround real-world problems. No need to resort to asymptotic theory as is usually the case when using frequentist methods in all but the simplest situations. For example, Bayesian probability bounds can be placed on almost any reliability quantity of interest as long as it is a function of parameters for which you can obtain posterior Markov chain Monte Carlo (MCMC) draws. Second, computational models such as equation-of-state and finite element models are universally used in all branches of science and engineering today.

When estimating reliability, frequentist methods ignore the information produced from such models and consider only traditional statistical data. In contrast, Bayesian methods utilize ALL available information when estimating reliability. It doesn't take a lot of brainpower to see that Bayesian methods have tremendous popular appeal among scientists and engineers. Because Bayesian methods exist on a level where they live, they will become even more widely used. Bayesian methods are obviously important in the new and exciting area of data fusion, also known as knowledge or information integration.

**Kvam:** Do you think there is a big future in this area for new Ph.D. researchers who are just starting out?

**Martz:** Certainly, but just how big a future is hard to say. Bayesian reliability methods couple nicely with new methods of optimization in high-dimensional spaces such as genetic algorithms. Together these new methods will result in research interest in highly optimized applications of Bayesian methods, particularly for complex systems. There will, of course, be a lot of interest in developing Bayesian networks in reliability areas. Reliability has always been a rich area for probabilistic modeling, such as renewal and repair. Subtle real-world nuances can produce nice theoretical models.

**Kvam:** You were awarded a Los Alamos National Laboratory Achievement Award, 1998, for pioneering work in the use of Bayesian statistical methods in health physics. What was your contribution?

**Martz:** Health physicists are interested in methods for determining whether or not a measured bioassay,

dosimetry, or other result should be called "positive," or "zero" (no detectable activity). Classical methods typically have unacceptably large false positive rates. I had the idea that prior knowledge regarding the distribution of true positives in the population should be useful for reducing this rate. If, for example, a facility historically has had a very small number of real positive results, then a new measured result should have to be a greater number of standard deviations away from zero to conclude that the new result is actually positive.

More recently, I suggested that MCMC could be used to solve the inverse problem of internal dosimetry: namely, how to use the bioassay measurements made on an individual over time to infer if and when intakes may have occurred and to estimate the magnitude of the resultant radiation dose. MCMC can do the job, but it takes a lot of computer time, even on a fast machine.

**Kvam:** How could you convince someone to be a Bayesian?

**Martz:** It takes a certain amount of statistical savvy and experience to fully appreciate the sad state of frequentist statistics and the beauty of Bayesian methods. For example, look at the convoluted interpretation that is attached to confidence intervals. In contrast, Bayesian probability intervals can be directly interpreted as a probability statement about a parameter, which is precisely what a practitioner desires. A prospective Bayesian must appreciate the value of all types of relevant information in a statistical analysis. This relevant information may arise from such sources as physical/chemical theory, computational results, generic industry-wide reliability data, past experience with similar devices, previous test results from a process development program and the subjective judgment of experienced personnel. Bayesian methods make appropriate use of all relevant data to supplement actual test data. This use of related information is a powerful advance; therefore, Bayesian methods have far greater appeal to scientists and engineers than classical methods. Finally, once someone has solved just a few problems using current MCMC software, such as WinBUGS, he or she will quickly become a Bayesian convert.

**Kvam:** We have known each other for 15 years, and I know faith is very important to you. How has it affected your career?

**Martz:** My faith in Jesus Christ has given balance to my life. It has put my priorities in order. After being "born again" in 1971, I placed my career where it should be: God, family, and then career. I stopped being so preoccupied with myself and my career. For example, I stopped working on evenings and weekends. I began to get involved more with helping others and am now a Gideon as well as being involved in church activities such as a local prison ministry. It's interesting how God has continued to bless my career even though I am less directly concerned with it. I believe that it was Dr. James Dobson who said, and I agree, that at the end of life there are only three questions that will be important: Whom did I love? Who loved me? What did I do for God in my life? I'm busy preparing answers to these questions.

**Kvam:** On the heels of the Challenger disaster, you published a paper on the risk of catastrophic failure on the space shuttle in *The American Statistician* (Martz and Zimmer, 1992). What led to that?

**Martz:** I knew and worked with Dr. Ben Buchbinder in the early 1980s when he was at the NRC. He then moved to NASA headquarters prior to the Challenger disaster on January 28, 1986. A few years later, Ben visited Los Alamos and talked with me about doing an analysis of some related data that he had collected regarding this disaster and other catastrophic military solid rocket booster failures. He referred me to the Marshall Space Flight Center for help in obtaining some other information that I needed. At the time, I was collaborating with Dr. Bill Zimmer, a statistician at the University of New Mexico, on some related empirical Bayes ideas, and I discussed Ben's request with him. Bill had some good ideas and, together, we decided to pursue this problem.

**Kvam:** Los Alamos has had its own catastrophes: the wildfire in 2000, the spy scandal occurring shortly after that (Wen Ho Lee was acquitted of charges), and infamous security breaches since then. How did these events affect you and other researchers in the lab?

**Martz:** It has been a traumatic time for all of us here. As a consequence, morale at the Lab has been on a rollercoaster ride for several years. Despite this, our statistics group is growing and the outlook is quite good. However, all of these things have affected me far less because I'm retired.

**Kvam:** It's been a long time since your Bayesian reliability book came out. Are you thinking of ever publishing another book?

**Martz:** Funny that you should ask. Val Johnson, Mike Hamada, Shane Reese, Alyson Wilson and I are just putting the finishing touches on a modern

MCMC-based sequel to be published by Springer-Verlag.

## 5. RETIREMENT

**Kvam:** You officially retired in 2001, but I notice you are still working and researching at the Laboratory. Do you have any plans on quitting soon?

**Martz:** I've been a Laboratory Associate since my retirement, and, in this capacity, I'm limited to working an average of only two days per week. This arrangement permits me to pretty much do only those things I like to do, such as mentoring and writing. Next year the Laboratory will be operating under a new contract, and I don't know if this program or something similar will continue. Provided it does, I would like to continue working part-time as long as I can contribute positively to the Lab and to the group.

**Kvam:** What projects keep your interest in research since your retirement?

**Martz:** Since I retired, I have become interested in several things. This past year I participated in a large knowledge management pilot project to develop a Bayesian network for an important application at the U.S. Strategic Command (STRATCOM). This project had many novel aspects that required research in Bayesian networks, which was fun. Hopefully, this project will continue. I also did some work with Mike Hamada on the combination of Bayesian accelerated and assurance testing, which we ultimately intend to submit for publication. Finally, I have been actively working on our forthcoming book on Bayesian reliability, which has also been great fun. Just enough involvement to keep me active but not overworked.

**Kvam:** What do you and Carol Ann have planned for the "real" retirement years?

**Martz:** Carol Ann also retired from the Laboratory in 2001 but continues to work part-time just as I do. We like to travel and, since retiring, have visited several new places, met lots of interesting people, and seen many wonderful things. We intend, God willing, to be able to do even more traveling. A few years ago we also purchased a vacation home in the mountain community of Angel Fire, New Mexico—about 90 miles from Los Alamos near the Colorado border. We're up there about one-third of our time now, relaxing and occasionally entertaining family and friends. Up north we also enjoy hiking, biking, snowshoeing, cross-country skiing, and, for me,

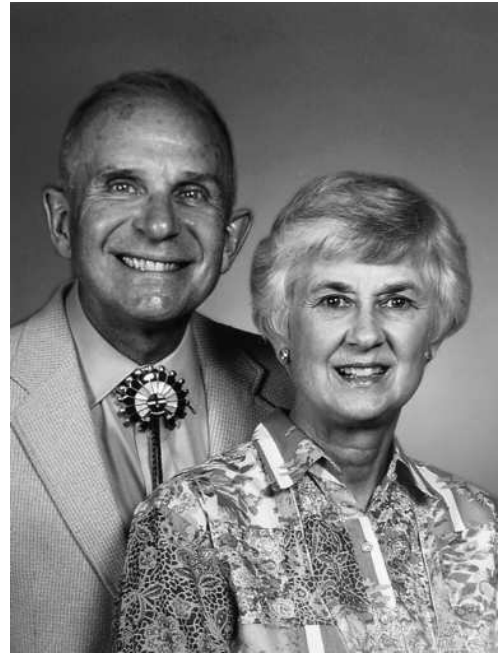

Fig. 2. *Harry and his wife Carol Ann in 2005.*

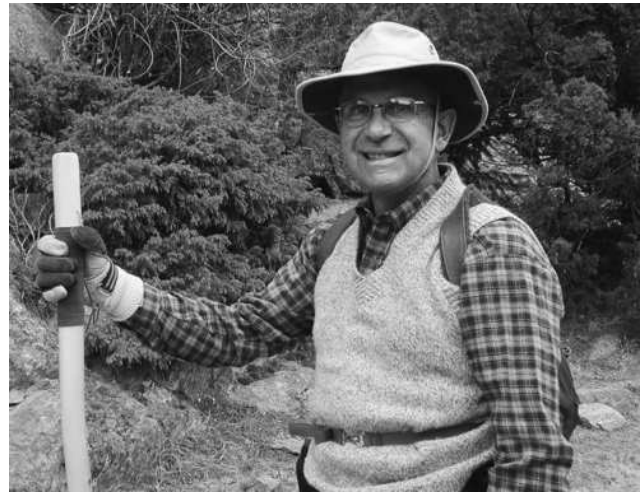

Fig. 3. *Harry hiking in Rocky Mountain National Park, CO in 2002.*

even motorcycling. Once we completely retire, it is likely that we will spend even more time there doing what retirees do—a little bit of this and a little bit of that. Both Carol Ann and I will probably volunteer for more church and community needs and activities then, as well. We really don't know how long we will maintain a home in Los Alamos once we completely retire. We have two children and two beautiful grandchildren living there now, and the answer partially depends on their future plans. I am also enjoying section hiking the Appalachian

Trail. Although I have completed over 700 miles of it so far, I still have about 1,400 more miles to go.

**Kvam:** Harry, I've thoroughly enjoyed chatting with you about your career and your life growing up. Thanks very much!